\documentclass{ws-ijmpe}

\begin{document}

\markboth{P. D. Stevenson, M. R. Strayer, J. Rikovska Stone and
  W. G. Newton}{Giant Resonances from TDHF}

%
\catchline{}{}{}{}{}
%

\title{GIANT RESONANCES FROM TDHF}

\author{\footnotesize P D STEVENSON}

\address{Department of Physics, University of Surrey\\
Guildford, GU2 7XH, United Kingdom\\
p.stevenson@surrey.ac.uk}

\author{\footnotesize M R STRAYER}

\address{Physics Division, Oak Ridge National Laboratory\\
Oak Ridge, TN 37831, United States of America}

\author{\footnotesize J RIKOVSKA STONE and W G NEWTON}

\address{Clarendon Laboratory, Department of Physics, University of Oxford\\
Oxford, OX1 3PU, United Kingdom}

\maketitle

\begin{history}
\received{(received date)}
\revised{(revised date)}
\end{history}

\begin{abstract}
A method of calculating giant resonance strength functions using
Time-Dependent Hartree-Fock techniques is described.  An application
to isoscalar giant monopole resonances in spherical nuclei is made,
thus allowing a comparison between independent 1-, 2- and
3-Dimensional computer codes.  
\end{abstract}

\section{Theory of Calculating Giant Resonances}
It is well-known that the Random Phase Approximation (RPA) is
equivalent to Time-Dependent Hartree-Fock (TDHF) to first order in the
density fluctuations from the static Hartree-Fock ground state.\cite{Row70}
Therefore either approach may be used to calculate giant resonance
states in nuclei.  Historically, the RPA approach has been more
popular, particularly working directly in the response function
formalism.\cite{Ber73}  This has especially been the case because of
computational complexity of TDHF calculations. Nevertheless,
successful calculations of giant resonances using TDHF have been made,
including those of Stringari and Vautherin,\cite{Str79} and Chomaz
{\it et al.}\cite{Cho87}

Recently there has been renewed interest in TDHF calculations of giant
resonances,\cite{Nak02}
motivated by the better scaling to calculations with no assumptions on
the spatial symmetry of the system.  In such cases, it is the TDHF
calculations which afford easier computation.  Such relaxation of the
symmetry assumptions are necessary for the proper calculation of
deformed systems and resonances.  It is also advantageous to use TDHF
if one wishes to explore different effective interactions, such as
finite range forces.  Once one has a static HF code it is, at least
conceptually, trivial to program a TDHF code.  Also, since RPA is a
limiting case of TDHF,  TDHF makes a natural choice for exploring
beyond-RPA (nonlinear) approaches.

To calculate giant resonance states, the TDHF equations
\begin{equation}
[\hat{h}(t),\rho(t)] = [\hat{h}_\mathrm{HF}[\rho(t)] + \hat{h}_\mathrm{ext}(t),\rho(t)] = i\hbar\dot{\rho}
\end{equation}
are solved.  This is achieved by solving the static HF equations
\begin{equation}
[\hat{h}_\mathrm{HF},\rho] = 0
\end{equation}
for the initial single-particle wavefunction $\{\psi_i(t=0)\}$.  These
are then evolved in time under the action of the unitary operator
\begin{equation}
U(t,t+\Delta t) = e^{-i\hat{h}\Delta t/\hbar}
\end{equation}
which is realised in the TDHF code by a Taylor expansion.  The
external perturbation used in the TDHF part consists of a spatial part
with a time profile, which couples to the density
\begin{equation}
\hat{h}_\mathrm{ext}(t) = \int\mathrm{d^3r}\,\rho(r,t)F({r})f(t)
\end{equation}
Here, $F(r)$ is the spatial form of the external perturbation,
which determines the kind of resonance which gets excited.  For the
present purposes isoscalar monopole (ISGMR) excitations are considered
and $F(r) = W_0 \sum_ir_i^2$, i.e. a harmonic oscillator potential
acting on all particles.  The function $f(t)$ describes the time
profile of the external perturbation, which is taken to be gaussian
in the following calculations.

The physically interesting observable is the strength function,
defined as
\begin{eqnarray}
S(E) &=& \sum_\nu\left|\langle\nu|F|0\rangle\right|^2 \nonumber \\
&=& -\frac{1}{\pi}Im\langle0|F\frac{1}{\hat{h}-E+i\delta}F|0\rangle
\end{eqnarray}
which is extracted from the TDHF calculation as the Fourier transform
of the fluctuation of the expectation of the spatial operator inducing
the excitation divided by the Fourier transform of the time profile of
the external perturbation.
\begin{equation}
S(\omega) = -\frac{1}{\pi}Im\int\mathrm{d^3r}\frac{\delta\langle
  F(r,\omega)\rangle}{f(\omega)}
\end{equation}

\section{Practical Calculations}
There are several issues to address when using TDHF to make practical
calculations.  Perhaps the most important involves boundary
conditions.  With TDHF calculations in coordinate space, one
necessarily works in a box of finite size.  If one imposes the
condition that wavefunctions vanish at the edge of the box then one
obtains a discrete excitation spectrum caused by particle flux
unphysically bouncing off the boundary and interfering with the
outgoing flux.  In this case, the number of discrete states found
depends upon the size of the box.  

One remedy is to use a box so large
that even after evolving for a long time,  flux has yet to reach the
boundary.  Even seventeen years ago, a box size of 720fm, with a rather
fine mesh (0.2fm)  was feasible for a spherically-symmetric (1D) TDHF
calculation of a resonance state.\cite{Cho87}  Nowadays, much larger
boxes can be easily used, at least in 1D.  One can think of this kind
of calculation where the maximum time evolved ($t_\mathrm{max}$) 
is sufficiently low that the boundary is not explored to be equivalent
to a continuum calculation.  Of course, the value of $t_\mathrm{max}$
determines the energy resolution of the calculated strength function.

For axially-symmetric (2-D) or triaxial (3-D) codes,  with which one can
look at more general kinds of resonance,  a smaller box size is
desirable for timely computation.  The discretisation that results in
the response function is unphysical, yet the discrete peaks lie in the
correct region, and the total (integrated) strength agrees with the
continuum result.  One may therefore perform a smoothing procedure on
the discretised results,  and compare to the smoothed continuum
result.  The smoothing process is physically reasonable when one
considers the experimental resolution of giant resonance strength and
the effect of spreading caused by higher-order effects not included in
RPA.  In this work a gaussian convolution is employed to smooth the
strength function.  

\begin{figure}
\centerline{\includegraphics[width=6.8cm]{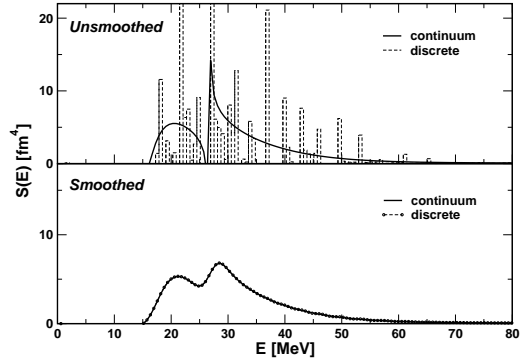}}
\caption{Strength function for Isoscalar Giant Monopole Resonance in $^{16}$O with BKN-like
  force. Without smoothing (upper panel), discrete and continuum
  calculations do not agree.  After smoothing (lower panel), agreement
  is excellent.  For further details, see text.\label{fig:en}}
\end{figure}

In figure \ref{fig:en}, the comparison is made between a continuum and
discrete calculation.  The continuum calculation is made in a 1000fm
box, and the discrete calculation in a 45fm box.  These are ISGMR
calculations in 1-D using a BKN-like force.\cite{Wu99}  In both cases
time evolution proceeded in steps of $\Delta t = 1$ Mev/c up to a
maximum time of $t_\mathrm{max}=2^{11}$ MeV/c.  The width of the
smoothing gaussian is 1MeV.  The unsmoothed discrete result is shown
as a histogram, which reveals the energy resolution commensurate with
the chosen value of $t_\mathrm{max}$.

\begin{figure}
\centerline{\includegraphics[angle=270,width=6.8cm]{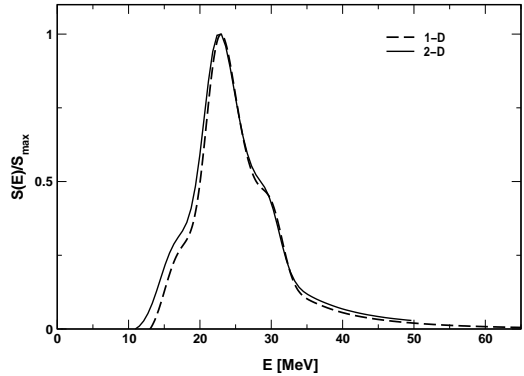}}
\caption{\label{fig:to} Isoscalar Giant Monopole resonance strength
  function using SkM* Skyrme interaction for $^{16}$O.  The 1-D
  calculations   corresponds to spherical symmetry, and the 2-D to
  axial symmetry.  The 1-D calculation was performed in a 24fm radius
  spherical box, and the   2-D calculation in a cylinder of height
  24fm and radius 24fm.  Gaussian smoothing is used with a width of 1MeV.}
\end{figure}
With higher-dimensional TDHF calculations, one is not restricted to
monopole resonances.  However, since it is difficult to perform true
continuum calculations in this case,  it is instructive to compare the
results of a monopole calculation from a code which allows deformation
to a spherically symmetric calculation.  This is a way to validate the
independent 1-,  2- and 3-D codes.  This is presented in figure
\ref{fig:to}.  In this case a more realistic Skyrme force,
SkM*,\cite{skms} is used, also to calculate the ISGMR in $^{16}$O.
The close agreement between the 1-D and 2-D calculations is
encouraging and suggests that the approach of using modest box sizes
along with smoothing is a feasible technique for calculating giant
resonances with TDHF. 


\vfill
\section*{Acknowledgements}
{\footnotesize  
This research was sponsored by the UK EPSRC, the US
  DOE grant no. DE-FG02-94ER40834 Division of Nuclear Physics,
  U.S. Dept. of Energy under contract DE-AC05-00OR 22725 managed by
  UT--Battelle. Support was also received from the Bergen
  Computational Physics   Laboratory in the framework of the European
  Community - Access to   Research Infrastructure action of the
  Improving Human Potential 
  Programme.  The authors acknowledge useful discussions with
  D. M. Brink and P.-G. Reinhard. }

\end{document}